\documentclass{blois}
\usepackage{amsmath,amssymb,amsfonts}
\usepackage{xcolor}

\def\be{\begin{equation}}
\def\ee{\end{equation}}
\def\bea{\begin{eqnarray}}
\def\eea{\end{eqnarray}}

\newcommand{\Photo}{}

\begin{document}

\vspace*{2cm}
\title{Higgs and Flavour: BSM Overview}

\author{ Joe Davighi }

\address{Theoretical Physics Department, CERN \\ 1211 Geneva 23, Switzerland}

\maketitle
\abstracts{
We discuss scenarios for BSM physics near the TeV, motivated by the hierarchy problem and the flavour puzzle, 
and review their experimental tests at present and future colliders. 
Strong LHC constraints on couplings to light quarks motivate $U(2)$-like flavour symmetries as a means of lowering the new physics scale: 
this is demonstrated by general SMEFT analyses, and is also seen in  
composite Higgs solutions to the hierarchy problem.
We discuss flavour non-universal gauge interactions as a possible origin for $U(2)$-like flavour symmetries which, in addition to allowing new physics to be lighter, opens up a simultaneous low-scale solution to the flavour puzzle. We focus on `flavour deconstructed' gauge interactions close to the TeV, and show how this non-universal gauge structure can be combined with Higgs compositeness in a way that better accommodates the requisite tuning in the Higgs mass.}

\section{A Higgs'-eye view of BSM}

Setting aside the cosmological constant, the Higgs is the origin of all fine-tuning puzzles in the SM: the hierarchy problem, the flavour puzzle, and the strong CP problem
(which is unphysical for massless quarks). 
Of these, only the hierarchy problem points to a particular mass scale for its solution, which is roughly that probed at high-energy colliders.
This means the Higgs plays a central role in shaping our pursuit of BSM physics at the high-energy frontier, both theoretically and experimentally. 
In this talk we focus on the Higgs being the origin of the hierarchy problem and the flavour puzzle. We discuss approaches to solving these two problems, possibly together, and review the experimental constraints and future prospects for testing such theories. 


The SM interactions involving the Higgs doublet $H$ are
\begin{equation} \label{eq:L}
    \mathcal{L} = |D_\mu H|^2-V(H) + \sum_{\psi\in u,d,e} y^\psi_{ij} \overline{\psi}_{L,i} H^{(c)} \psi_{R,j}, \quad V(H) = -\mu^2 |H|^2+\lambda |H|^4\, .
\end{equation}
These interactions reveal two structural puzzles in the SM.
The {\em hierarchy problem} is that, regarding the SM as an effective field theory with a cut-off beyond the TeV, the dimensionful parameter $\mu^2$ is unnaturally small. If there exist any heavy BSM particles $X$ with masses $M_X$ that have been integrated out ({\em e.g.} for neutrino masses, grand unification, 
or even as a stand-in for quantum gravity effects), then we expect $n$-loop threshold corrections to $m_h^2=2\mu^2$ that scale as $\delta m_h^2 \sim (16 \pi^2)^{-n} g^{2n} M_X^2$, unless there is some mechanism to soften this quadratic sensitivity. 

The majority of the Higgs' couplings, however, are generating flavour. These Yukawa couplings $y^{u,d,e}_{ij}$ also display a highly non-generic structure;
the top Yukawa $y_t$ is order-1 in these units, while the other 12 physical parameters are much smaller and strongly hierarchical. 
This unexplained structure constitutes the {\em flavour puzzle}, whose solution likely entails flavour non-universal dynamics. 
While the hierarchy problem points to a mass scale, the Yukawa couplings depend only logarithmically on heavy mass scales, meaning the flavour puzzle could in principle be decoupled from the hierarchy problem and solved deep in the UV.


A well-known solution to the hierarchy problem is to suppose the Higgs is not elementary but composite, in which case loop corrections to $m_h^2$ are cut-off by the compositeness scale. 
To get $m_h \ll M_\star$, where $M_\star $ denotes the masses of other composite resonances that continue to elude LHC searches, the composite Higgs (CH) should moreover be a pNGB associated with some global symmetry breaking $G\to H$, where $H$ contains custodial symmetry. 
Explicit breaking of $G$ by $y_t$ and by gauging $SU(2)_L \times U(1)_Y$ generates the Higgs potential at 1-loop, with the parametric dependence
    $\delta m_h^2|_{\mathrm{CH}} \sim \frac{1}{16\pi^2} (\alpha y_t^2 M_T^2-\beta g^2 M_\rho^2)$, for model-dependent coefficients $\alpha,\beta\in\mathbb{R}_+$, and where $M_{T,\rho}$ are the masses of the lightest composite spin-1/2 and spin-1 resonances.
Another well-known solution to the hierarchy problem is supersymmetry, wherein the inclusion of loops involving superpartner particles softens the quadratic sensitivity of $\delta m_h^2$ to the UV cut-off to a logarithmic sensitivity, {\em viz.} $\delta m_h^2 \sim \frac{1}{16\pi^2} y_t^2 M_T^2 \log (\Lambda^2/M_T^2)$ where again $M_T$ is the top partner mass (now a stop). 
In either the case, delivering the observed Higgs mass without fine-tuned cancellations roughly predicts heavy particles with masses $M_\star \lesssim 4\pi m_h \sim$ few TeV.

But while the hierarchy problem points to a scale $M_\star \sim$ TeV, precise measurements in flavour physics point to much higher scales~\cite{EuropeanStrategyforParticlePhysicsPreparatoryGroup:2019qin}. For example, dimension-6 flavour-violating effective operators of the form $\mathcal{L}\supset (\overline{d}s)^2/\Lambda_{sd}^2$, which would arise from integrating out a heavy scalar or vector with flavour-violating couplings to the SM, must be suppressed by an effective scale $\Lambda_{sd}\gtrsim 10^{5\div 6}$ TeV to not contravene bounds from kaon mixing. 
This and other bounds tell us that {\em any} solution to the hierarchy problem needs a non-trivial flavour structure that somewhat resembles that of the SM. Minimal flavour violation~\cite{DAmbrosio:2002vsn} (MFV) is one such structure, in which 
the SM Yukawa couplings $y^\psi$ are the only source of flavour violation in the full theory, and in the limit $y^\psi\to 0$ the full Lagrangian respects the large $U(3)^5$  global symmetry that the SM sector would then possess.
For instance, the couplings $C^X_{ij}$ of a SM gauge singlet $X$ to SM fermion bilinears $\overline{\psi}_i \psi_j$ would take the form $C_{ij}^{\mathrm{MFV}}=\delta_{ij}+\tilde{c}_{ij}(y^{u,d,e})$ for some functions $\tilde c$. 
For kaon mixing we can see that MFV gives enough suppression to bring us down to the TeV, predicting $\Lambda_{sd}^{-2}\sim y_t^4 (V_{31}V_{32}^\ast)^2 \Lambda_{\mathrm{high}}^{-2} \sim (10^5 \Lambda_{\mathrm{NP}})^{-2}$. 

The scale  $M_{\star}$ has been probed directly through high-$p_T$ searches at ATLAS and CMS, which  set limits of a few TeV on {\em e.g.} stop squarks~\cite{ATLAS:2024lda} and composite top partners~\cite{ATLAS:2023bfh}.
For SUSY or CH with MFV, some fine-tuning is thence required, of order $\delta m_h^2/m_h^2 \sim (M_T / 500 \mathrm{~GeV})^2$. Nor is there any sign of compositeness in measurements of the Higgs' couplings: $HWW$ and $HZZ$, in particular, are predicted to be shifted by a universal amount in CH models, and their measured agreement~\cite{ATLAS:2020qdt,CMS:2020gsy} with the SM to within about 3\% dictates that $v^2/f^2 \lesssim 5 \%$ or so.

Thus, the current state of play is that our most well-developed solutions to the (large) hierarchy problem suffer from a `little hierarchy': an unavoidable tuning between different TeV$^2$-sized contributions to $\delta m_h^2$ to within a few percent. 
One path forward is to accept some tuning and proceed undeterred in our search for SUSY or compositeness. Deviating slightly, one might try smarter model-building within either of these paradigms to offer a symmetry-based account of the little hierarchy~\cite{Durieux:2021riy}.
Flavour provides us with useful model-building tools in this regard. By going beyond MFV and considering flavour {\em non}-universal dynamics, while still respecting $U(2)^5$ symmetries that sufficiently suppress {\em e.g.} kaon mixing shifts,
one can reduce the size of the little hierarchy appreciably.
We discuss this in the context of CH models~\cite{Glioti:2024hye}, which corroborates conclusions obtained using the SM effective field theory (SMEFT). 
The fact that non-universal dynamics can facilitate more natural solutions to the hierarchy problem suggests the intriguing possibility that the flavour puzzle might also be solved at low-scales, tied to the hierarchy problem. We illustrate some recent model-building sketches in this direction, which use the idea of `flavour deconstruction'.

\section{Flavour symmetries: from MFV to flavour non-universal New Physics} \label{sec:MFVtoU2}

The agreement of LHC results with SM predictions puts pressure on TeV-scale new physics models with MFV, with strong constraints coming from the 
unsuppressed couplings to valence quarks which implies unsuppressed production at $pp$ colliders {\em e.g.} via Drell--Yan. 
While the strength of bounds is model-independent, as an example, LHC searches for heavy replicas of the (flavour-universal) SM $Z$ or $W$ bosons put bounds of 5 TeV and 6 TeV respectively~\cite{ParticleDataGroup:2024cfk}.

The desire to realise lighter new physics, and so facilitate more natural solutions to the hierarchy problem, motivates other flavour structures. 
In pursuit of alternatives, it is instructive to revisit the approximate global symmetries of the SM. While the SM Yukawa couplings $y^{u,d,e}_{ij}$ 
break the $U(3)^5$ flavour symmetry of the SM kinetic and gauge terms down to $U(1)_B \times \prod_{i=1}^3 U(1)_{L_i}$, this is only a {\em weak breaking} because most of the Yukawa couplings are small. Keeping only $y_t\neq 0$ leaves an unbroken $U(2)_q \times U(2)_u \times U(3)^3$, with light quarks transforming as $U(2)$ doublets and the third family as $U(2)$ singlets. The leading spurions that further break these symmetries are those generating $V_{cb}\sim 0.04$ and $y_c/y_t\sim 0.01$.
If new physics respects these smaller $U(2)^n$ symmetries, one achieves just as good protection of the most sensitive flavour observables~\cite{Barbieri:2011ci,Isidori:2012ts}.

There are two important benefits to passing from MFV to $U(2)$-like flavour structures for new physics. The first is theoretical, which is that $U(2)^n$ mimics the flavour structure of SM Yukawas, by design. This means that, while we lose the predictive power of MFV (whereby $Y^\psi$ are input as spurions), we open the possibility that the same dynamics explaining the $U(2)$ structure of BSM couplings {\em also} explains the SM flavour puzzle.
The second benefit is phenomenological: in passing from MFV to $U(2)$ one can realise new physics at lighter scales, as anticipated. With $U(2)$, the BSM couplings to SM fermion bilinears $\overline{\psi}_i \psi_j$ can now take the {\em non}-universal form $C_{ij}^{U(2)}=\mathrm{diag}(a,a,b)$, with $b\neq a$. By going to the limit $b \gg a$, one can largely decouple such BSM particles from light quarks and so weaken many otherwise strong bounds, as we illustrate in \S\S \ref{sec:SMEFT} and \ref{sec:CH}.

\subsection{Lessons from the SMEFT} \label{sec:SMEFT}

High $p_T$ searches in ATLAS and CMS in $pp \to ll$ put strong constraints on particles produced via Drell--Yan $q\bar{q}\to X \to ll$; the strength of the bound is significantly weaker if the BSM particle couples only to third generation quarks, due to the suppression of bottom quark PDFs in the proton.
If the BSM particle mass is beyond the kinematic range of the search ($\sim 3$ TeV), one can integrate out the heavy particle to obtain semi-leptonic dimension-6 SMEFT operators, {\em e.g.} $\mathcal{L}\supset \frac{1}{\Lambda^2}[C_{lq}^{(1)}]_{aaij} (\overline{l}_{a}\gamma^\mu l_a) (\overline{q}_i \gamma_\mu q_j)$.
For $i=j=1$ and $a\in\{e,\mu\}$, current bounds reach $\Lambda/\sqrt{C} \gtrsim 10$ TeV~\cite{Allwicher:2022gkm}, whereas for heavy quark flavours $i=j=3$ this drops to $\Lambda/\sqrt{C} \gtrsim 1.5$ TeV.

This lowering of the new physics scale by aligning with the third generation is not specific to Drell--Yan tails. At the level of SMEFT operators, it has been demonstrated to hold globally across data from flavour, electroweak (including $Z$-pole, $W$-pole, flavour-conserving Higgs decays, and tests of lepton flavour universality in tau decays), and colliders (including also jet observables)~\cite{Allwicher:2023shc}. 
In particular, it has been shown that if operators involving $n$ light quark fields are suppressed by a factor $(0.16)^n$ relative to operators involving only third-generation quarks, then the sensitivity across all experimental data (as quantified via a combined likelihood) drops to the 1.5 TeV level, that is the sensitivity to pure third generation operators from LHC data. This modest degree of third-family alignment is natural in models with $U(2)$ flavour symmetry, given the largest spurion is that generating $V_{cb}$ which is a doublet under $U(2)_{q_L}$ with magnitude set by $|V_{cb}|\sim 0.04$.

It was also demonstrated~\cite{Allwicher:2023shc} that the proposed `tera-$Z$' run at FCC-ee would bring a spectacular improvement in sensitivity to new physics aligned with the third generation, due to the unavoidable 1-loop running of most SMEFT operators into observables measured on the $Z$-pole. Including only the projected improvements in $Z$-pole and $W$-pole observables from FCC-ee, the sensitivity to the $U(2)$-like SMEFT scenario described above would jump from 1.5 TeV to $5\div 10$ TeV.

\subsection{Lessons from the Composite Higgs} \label{sec:CH}

The SMEFT studies discussed in the previous Subsection put bounds on a single effective operator at a time, and so cannot capture any of the correlations that would occur in a concrete BSM model. 
Since the central motivation for new physics near the TeV is the hierarchy problem, it is important to address whether the apparent advantage brought by $U(2)$ over MFV in lowering the scale is actually realised in solutions to the hierarchy problem.

Different flavour symmetries in the context of CH models have been revisited in recent works~\cite{Glioti:2024hye,Stefanek:2024kds}. In order to even broach the subject of flavour, one must first recap how Yukawa couplings can be generated in a CH model. Na\"ively coupling bilinears of elementary SM fermions to some composite scalar operator $\mathcal{O}_H$ is problematic. One needs the quantum dimension $d_\mathcal{O}$ of $\mathcal{O}_H$ to satisfy $d_\mathcal{O}\approx 1$ to get a large enough top Yukawa, given the UV cut-off scale $\Lambda$ cannot be low due to flavour bounds, while at the same time we want the operator $\mathcal{O}_H \mathcal{O}_H^\dagger$, whose dimension should be $d_{\mathcal{O}\mathcal{O}}\approx 2d_\mathcal{O}\approx 2$, to be irrelevant to not reintroduce a hierarchy problem. Partial compositeness~\cite{Kaplan:1991dc} (PC) provides a solution, in which SM fermions mix {\em linearly} with composite fermionic operators, {\em viz.} $\mathcal{L}_{\mathrm{PC}}\supset \lambda^{ia}\psi_i \mathcal{O}_a$, which then mix with the composite scalar operator, {\em viz.} $\mathcal{L}_{\mathrm{PC}}\supset \overline{\mathcal{O}}_a \mathcal{O}_H \mathcal{O}_b$. The PC idea even  promised a {\em dynamical} solution to the SM flavour puzzle, whereby exponential hierarchies would be generated starting from anarchic order-1 couplings $\lambda^{ia}$ 
purely by RG running 
of the mixing operators from $\Lambda$ down to $M_\star$, given some small differences in the anomalous dimensions %
of the composite operators $\mathcal{O}_a$. However, this anarchic scenario unavoidably entails also large flavour violation at the scale $M_\star$, and so cannot be realised below $M_\star \sim \mathcal{O}(100\mathrm{~TeV})$, even assuming an optimistic scenario in which electric dipole moment corrections are 1-loop suppressed~\cite{Liu:2016idz}. 
This motivates the use of a flavour symmetry 
to bring down the scale $M_\star$. 

With an MFV-like flavour hypothesis, and assuming a custodial symmetry $(SU(2)_L \times SU(2)_R) \rtimes \mathbb{Z}_2$ to limit corrections to both the $m_W/m_Z$ ratio and the $Zb_Lb_L$ vertex, a comprehensive analysis of current constraints shows that $M_\star \gtrsim 7 \div 8$ TeV~\cite{Glioti:2024hye}. The most important constraints are driven by order-1 couplings to light generation fermions, namely di-jet $pp \to jj$ LHC measurements and modified $W$-couplings to light quarks, as well as precision flavour tests such as $B$-meson mixing and $BR(B_s \to \mu^+ \mu^-)$ measurements at the LHC. In other words, flavour provides the sharpest experimental tests of CH solutions to the hierarchy problem, not current bounds from electroweak precision observables (EWPOs).
With $U(2)$-like flavour symmetries, most of these constraints decouple apart from the $b\to s$ tests, allowing $M_\star \gtrsim 2$ TeV to be realised~\cite{Glioti:2024hye}, and hence a significant reduction in the little hierarchy tuning compared to the MFV-like scenarios. This corroborates the general lessons we learnt in \S \ref{sec:SMEFT} from the SMEFT.

Again as we learnt from SMEFT studies, the potential to push back indirect sensitivity to a CH-like scenario at FCC-ee should be significant, due to the unavoidable shifts in electroweak precision observables. It has been estimated, again using SMEFT as a tool but with a strongly-interacting light Higgs (and top) inspired EFT power counting, that with FCC-ee the sensitivity to $U(2)$-like CH models will jump from 2 TeV to more than 20 TeV~\cite{Stefanek:2024kds}. Even four-top operators give significant (2-loop) shifts to EWPOs~\cite{Allwicher:2023aql} {\em e.g.} the $W$-mass at the level of precision anticipated at FCC-ee. This suggests that, even with $U(2)$ flavour symmetry, a CH solution to the hierarchy problem cannot hide from FCC-ee without the requisite tuning jumping from $\mathcal{O}(10^{-2})$ to the $\mathcal{O}(10^{-4})$ level.
In the case that tera-$Z$ does reveal deviations indicative of compositeness,
it should be emphasized that a high-energy next generation collider would then have direct reach up to $M_\star$ of order 20 TeV for FCC-hh~\cite{Golling:2016gvc} or $E_{\mathrm{CoM}}/2$ for a Muon Collider~\cite{Accettura:2023ked}.

\section{From global symmetries to non-universal gauge interactions}

So far, approximate $U(2)^n$ or $U(3)^n$ flavour symmetries have been imposed, and their phenomenological consequences explored (both using SMEFT, and using CH solutions to the hierarchy problem as a case study.)
But what could be the UV origin of these flavour symmetries?
This question leads us to the other main benefit of passing from MFV to $U(2)$-like flavour structure, which is the opportunity to explain the SM flavour puzzle with the same $U(2)$-preserving BSM dynamics.

\subsection{Horizontal gauge models}

A simple route to obtaining global $U(2)$ flavour symmetries controlling both SM and BSM flavour textures is if the global symmetry is obtained {\em accidentally}, meaning it is a consequence of a flavour non-universal {\em gauge} symmetry under which the first two generations are charged equally, but the third generation is charged differently.
But what symmetry should one gauge? There are of course many options. One approach, in the spirit of Froggatt--Nielsen~\cite{Froggatt:1978nt}, is to directly gauge some subgroup $G_{\mathrm{hor}} \subset U(2)^n$ of the flavour symmetry, and break it to nothing. This will deliver a sector of $Z^\prime$ bosons that mediate direct flavour violation, so they cannot reside near the TeV scale. For example, if one gauges $SU(2)_{q+l}$ under which the left-handed light generation quarks and leptons transform as doublets~\cite{Greljo:2023bix}, the resulting $Z^\prime$ mediates the lepton-flavour-violating $K_L \to \mu^\pm e^\mp$ decay, which implies $M/g \gtrsim 10^{2\div 3}$ TeV. On the other hand, for such horizontal flavour models one is free to take the limit in which the new gauge coupling $g\to 0$, so at least one can decouple the loop contributions to the Higgs mass (which scale as $\delta m_h^2 \sim g^4 v^2/16\pi^2$) at the same time as taking $v \gg v_{\mathrm{EW}}$. It has recently been shown that this horizontal approach, based on gauging $SU(2)_{q+e}$, can elegantly reconcile hierarchical quark mixing with anarchic neutrino mixing~\cite{Antusch:2023shi}.

\subsection{Flavour deconstruction: solving the flavour puzzle near the TeV} \label{sec:deconstruction}

\begin{table}[t] \label{tab:survey}
\caption[]{Qualitative comparison of flavour deconstruction models~\cite{Davighi:2023iks}.
The quoted `natural upper limits' estimate the finite part of the leading radiative corrections to $m_h^2$, and require these be less than $1 ~\mathrm{TeV}^2$.}
\label{tab:exp}
\vspace{0.4cm}
\begin{center}
\begin{tabular}{|c|c|c|c|c|c|c|}
\hline
& Deconstructed force & $SU(3)$ & $SU(2)_L$ & $SU(2)_R$ & $U(1)_Y$ & $U(1)_{B-L}$ \\
\hline
Flavour & $|V_{cb}| \ll 1$ & $\checkmark$ & $\checkmark$ & $\times$ & $\checkmark$ & $\checkmark$ \\
 & $y_i\ll y_3$ & $\times$ & $\checkmark$ & $\checkmark$ & $\checkmark$ & ${\times}$ \\
\hline
EW & Natural upper limit of $|\tan\theta| M$ & 90 TeV & 20 TeV & 40 TeV & 40 TeV & 500 TeV \\
 & EWPOs order & 1-loop & Tree & Tree & Tree & 1-loop \\
\hline
\end{tabular}
\end{center}
\end{table}

An alternative approach, in which there is no direct flavour violation and which can therefore reside near the TeV, is based on {\em flavour deconstruction} of the SM gauge interactions~\cite{Li:1981nk}. Here, one splits a SM gauge interaction $G$ into three copies $G_i$, one that acts on each generation with the Higgs coupled to $G_3$; the phenomenology is determined by the 2-step breaking pattern
\begin{equation}
    G_1 \times G_2 \times G_{3+H} \xrightarrow{v_{12}\sim 10^{2\div 3} \mathrm{~TeV}}G_{12}\times G_{3+H} \xrightarrow{v_{23}\sim 1 \mathrm{~TeV}} G_{\mathrm{SM}}\, ,
\end{equation}
and so $U(2)$ symmetries hold up to scales $v_{12}$. This $\prod_i G_i$ gauge symmetry could arise, for instance, from a theory with unified electroweak and flavour symmetries~\cite{Davighi:2022fer,Davighi:2022vpl} or from an extra dimension~\cite{Fuentes-Martin:2022xnb}.

How does this solve the SM flavour puzzle? We suppose the breaking steps are triggered by the condensing of scalar fields $\phi_{12}$ and $\phi_{23}$ transforming in the bi-fundamental representation of $G_{1}\times G_{2}$ and $G_2\times G_{3+H}$ respectively.\footnote{
One reason why flavour deconstruction is appealing is that the breaking of $G_{A}\times G_B$ to its diagonal subgroup is generic; it does not depend on the representation of $\phi$, provided only it is charged under both $G_A$ and $G_B$, nor on the values of $g_{A,B}$. This is because there is no other non-trivial embedding of semi-simple $G$ into a group $\cong G \times G$.}
Because the Higgs is charged under $G_{3+H}$ only third-generation Yukawa couplings are gauge invariant; to form an effective Yukawa coupling $\sim y_{23}\overline{\psi}_{L,2}H \psi_{R,3}$ requires an insertion of $\phi_{23}$ and so is a higher-dimension operator, that could arise {\em e.g.} from integrating out a vector-like fermion $\Psi$ with mass $M_\Psi$. The resulting Yukawa coupling $y^{f}_{23}$ ends up suppressed by a ratio of scales $\epsilon_{23}:=v_{23}/M_\Psi$, which we expect to be of order $|V_{cb}|$. In this way, suitably hierarchical 3-generation Yukawa textures can be generated, depending on which SM gauge interaction(s) we deconstruct~\cite{Davighi:2023iks}.

This deconstruction approach to solving the flavour puzzle allows, and arguably predicts, richer phenomenology than the horizontal approach. The TeV breaking step delivers a set of heavy gauge bosons charged in the adjoint of $G$, with flavour diagonal {\em but non-universal} couplings to SM fermions that respect $U(2)$. In particular, $C_{ij}^{U(2)}\sim g_{\mathrm{SM}}\,\mathrm{diag}(\frac{g_{12}}{g_3}, \frac{g_{12}}{g_3}, \frac{g_3}{g_{12}})$, where the gauge couplings satisfy a matching condition $g_{12}^{-2}+g_3^{-2}=g_{\mathrm{SM}}^{-2}$ that puts a lower limit on the couplings of the heavy vectors, $g_{12,3}\geq g_{\mathrm{SM}}$. As a result, and in contrast to the horizontal case, one cannot decouple the flavour deconstruction dynamics to high scales by taking the gauge boson mass $M \sim gv \to \infty$ without creating a hierarchy problem, given the 1-loop Higgs mass correction $\delta m_h^2 \sim g^2 M^2/(16\pi^2)$. This means that, if we want to explain the flavour puzzle via deconstruction, we are forced to take the hierarchy problem seriously, and so we are led into realising the deconstruction dynamics at a low scale and/or solving the hierarchy problem at the same time.

\subsection{Phenomenology of deconstructed electroweak forces} \label{sec:pheno}

With this in mind, we report the phenomenology of deconstruction models that solve the flavour puzzle -- after, we will discuss how the hierarchy problem could be solved at the same time via compositeness. To chart the parameter space, we define a gauge mixing angle $\tan \theta :=g_{3}/g_{12}$ that quantifies the degree of alignment of the new physics with the third family (with third family alignment corresponding to $\theta \to \pi/2$); then observables can be computed as a function of $\theta$ and the mass $M$ of the gauge boson multiplet, given an assumption about CKM and lepton mixing. 

To solve the flavour puzzle necessarily involves deconstructing part of the electroweak force~\cite{Davighi:2023iks}, as is clear from Table~\ref{tab:survey}, with two immediate consequences: (i) there are tree-level shifts to EWPOs that give multi-TeV constraints, and (ii) there are unavoidable one-loop corrections to the Higgs mass.
For instance, if we compute the finite parts of the 1-loop $\delta m_h^2$ corrections and demand these be less than $1~\mathrm{TeV}^2$ to not worsen the little hierarchy, then $\tan\theta M \lesssim 20$ TeV in the case of deconstructed $SU(2)_L$, and $\tan\theta M \lesssim 40$ TeV for deconstructed hypercharge (the difference being due to $g_Y \approx g_L/2$). These limits define natural regions for these flavour models.

For both the cases of deconstructed $SU(2)_L$~\cite{Davighi:2023xqn,Capdevila:2024gki} and deconstructed $U(1)_Y$~\cite{Davighi:2023evx,FernandezNavarro:2023rhv},
current constraints already significantly eat into these natural regions, with EWPOs, flavour tests, and high-energy LHC measurements being highly complementary. A key message is that, given flavour deconstruction does {\em not} introduce new flavour violation (in contrast to gauging horizontal flavour symmetries) but {\em does} necessarily introduce heavy vectors coupled to the Higgs, the constraints on these flavour models coming from  EWPOs are often stronger than those coming from flavour. This complements what we saw for CH models in \S \ref{sec:CH}, that solutions to the hierarchy problem can receive their strongest constraints from flavour observables.
In the case of deconstructed $SU(2)_L$ the electroweak constraints are particularly strong, in part because the model predicts $m_W<m_W^{\mathrm{SM}}$, shifted in the opposite direction to the average of current $m_W$ measurements (including or excluding the CDF 2022 measurement); in contrast, deconstructing hypercharge shifts $m_W/m_Z$ in the positive direction, and so can even improve the quality of the EW fit with respect to the SM. 

For both models, measurements of $BR(B_s \to \mu^+ \mu^-)$ are the most stringent flavour tests, given (i) the impressive experimental precision achieved especially by LHCb and CMS, (ii) that this observable is theoretically clean, depending only on the axial $O_{10}\sim (\overline{b}\gamma_\mu P_L s)(\mu \gamma^\mu \gamma^5 \mu)$ operator, and (iii) the fact that both models predict a large shift in the corresponding $C_{10}$ coefficient. Constraints from $B_{d,s}$ meson mixing are also important, but strictly weaker than those from $B_s \to \mu\mu$, echoing what we saw in \S \ref{sec:CH} for CH models with $U(2)$-like flavour symmetry. Finally, as expected, high-energy LHC measurements in $pp \to ll$ give the strongest constraints -- which we computed using the \texttt{HighPT} tool~\cite{Allwicher:2022mcg} -- in the parameter space region with $g_{12} > g_3$, which pushes us toward the third-family aligned region (reflecting our general comparison of MFV {\em vs.} $U(2)$-like flavour structures in \S \ref{sec:MFVtoU2}.) We summarize the key constraints on both models in Table~\ref{tab:exp}.

\begin{table}[t]
\caption[]{Summary of constraints~\cite{Davighi:2023xqn,Davighi:2023evx} from flavour, high $p_T$, and EWPOs on the mass $M$ of the gauge bosons predicted by the breaking pattern $G_{12}\times G_{3+H}\to G_{\mathrm{SM}}$ for $G=SU(2)_L$ and $U(1)_Y$.}
\label{tab:exp}
\vspace{0.4cm}
\begin{center}
\begin{tabular}{|c|c|c|c|}
\hline
& Deconstructed $SU(2)_L$ & Deconstructed $U(1)_Y$ \\
\hline
Electroweak: $Z$-pole \& $W$-pole & 9 TeV (5 TeV if exc. $m_W$) & 2 TeV  \\
Flavour: $B_s\to\mu^+\mu^-$ (up-alignment) & 7.5 TeV & 2 TeV \\
High $p_T$: Drell--Yan $pp\to ee,\mu\mu,\tau\tau$ & 4.5 TeV & 3.5 TeV \\
\hline
EW projection FCC-ee & 30 TeV & 7 TeV  \\
\hline
\end{tabular}
\end{center}
\end{table}

Turning to future prospects, the importance of EWPOs in constraining these flavour models means a tera-$Z$ run at FCC-ee would be sensitive up to very high scales: we find that, if measurements are SM-like, masses below 30 TeV will be exlcuded for deconstructed $SU(2)_L$ (for all values of the mixing angle $\theta$), essentially covering the entire `natural' parameter space for which $\delta m_h^2 \lesssim (\mathrm{TeV})^2$. A similar, albeit slightly less dramatic, reach is expected for the deconstructed $U(1)_Y$ version. Before then, High-Luminosity LHC would already bring significant improvements in the medium term, both through increased precision on $B_s \to \mu \mu$ measurements but also through the great boost in statistics for the high $p_T$ Drell--Yan searches; for deconstructed $SU(2)_L$, the Drell--Yan reach should improve from 4.5 TeV to 8 TeV after HL-LHC.

Finally, it should be emphasized that FCC-ee is not just an electroweak precision machine; of particular relevance to theories of Higgs and flavour, FCC-ee will make great strides forward also in precision flavour measurements~\cite{Monteil:2021ith}. A flagship process is the prospect of a $B \to K^\ast \tau\tau$ measurement at the SM rate~\cite{Kamenik:2017ghi}, which would be highly sensitive to third-generation aligned BSM physics. Per-cent level precision in $bs\nu\nu$ processes~\cite{Amhis:2023mpj} is also feasible, which would be particularly powerful in combination with the HL-LHC precision on $bs\mu\mu$ processes given the two are directly correlated in these models. FCC-ee will also serve as a `tau factory', with tau LFUV measurements alone being able to probe the deconstructed $SU(2)_L$ scenario up to $M \geq 13$ TeV.

\section{Towards a Joint Approach to Higgs and Flavour}

We have seen that, on the one hand, CH solutions to the hierarchy problem require a $U(2)$-like flavour symmetry to realise a more natural scale $M_\star \approx 2$ TeV. On the other hand, we saw how such $U(2)$ symmetries can emerge accidentally from flavour deconstructed gauge interactions that, if there, prefer to be broken close to the TeV scale to avoid large $m_h^2$ corrections, but that  EWPOs (in particular) push us to regions with finite $\delta m_h^2$ corrections approaching the TeV$^2$.
This motivates us to explore a joint solution to the hierarchy problem and the flavour puzzle close to the TeV, and to assess how each of these two model-building hypotheses feeds into the other~\cite{Covone:2024elw}.

We postulate a strong sector that delivers the global symmetry breaking pattern of the minimal CH model~\cite{Agashe:2004rs}, but taking the unbroken custodial subgroup to be the flavour non-universal $H=SU(2)_L \times SU(2)_{R,3}$. Strong dynamics is presumed to break $Sp(4) \to H$ at strong coupling scale $\Lambda_{\mathrm{HC}}:=8\pi F$, delivering a PNGB Higgs transforming in the bidoublet of $SU(2)_L \times SU(2)_{R,3}$. At a scale $v_{23}$ the gauge group linking $U(1)_{R,3}\times U(1)_{B-L}\times U(1)_{Y,12} \to U(1)_Y$ occurs, due to the condensing of additional scalar fields that may be elementary or also composite. 
The gauged $U(1)_{B-L}\times U(1)_{Y,12}$ factor, together with the light generation SM fermions, are external to the composite dynamics. 
The pNGB Higgs couples to the third generation via the usual PC framework described in \S \ref{sec:CH}, via linear mixing with composite top partners, {\em viz.} $\mathcal{L}\supset \overline{t}_{L/R}\mathcal{O}$ {\em etc}. The Higgs couples to the lighter generations again via the composite top partners $\mathcal{O}$, but now the linear mixing between these and the light fermion fields requires insertion of the link field $\phi_{23}$ to be invariant under the deconstructed gauge symmetry; this could be resolved at shorter distances by a vector-like fermion $\Psi$ with interactions $\mathcal{L}\supset \overline{\Psi} \mathcal{O}+ \overline{\psi}_2\phi_{23}\Psi$ where $\psi_2$ is a second generation SM fermion. This mechanism generates hierarchical Yukawa couplings in the same fashion as for the elementary flavour deconstruction set out in \S \ref{sec:deconstruction}; at the same time, the deconstructed gauge symmetry enforces $U(2)$ accidental symmetries on all couplings between the strong sector and the SM, and so the phenomenology associated to the composite dynamics inherits the $U(2)$ protection that we reported in \S \ref{sec:CH} -- this global symmetry now has a simple dynamical explanation.

In addition to solving the SM and BSM flavour puzzles, deconstructing the CH model brings further benefits which can be seen by computing the 1-loop generated Higgs potential. 
Enforcing a left-right symmetry to eliminate a divergent contribution~\cite{Csaki:2017cep}, the
Higgs mass is~\cite{Covone:2024elw}
\begin{equation}
    m_h^2 = -\frac{1}{16\pi^2}\left[ 4N_c y_t^2 M_T^2 - \frac{9}{2}g_{R,3}^2 M_\rho^2\left( 1- \frac{2M_{W_R}^2}{M_\rho^2}\right) \right]
\end{equation}
Because the gauge coupling $g_{R,3} = g_{\mathrm{SM}}/\cos\theta$ can be pumped up by going to the third-family-aligned limit, which is independently favoured by LHC data (\S \ref{sec:pheno}), one can better tune the gauge {\em vs} top quark contributions to deliver $m_h \sim 100$ GeV. Numerically, this allows for the top partner to be heavier ($M_T > 2$ TeV), affording better compatibility with direct searches than is possible with the usual minimal CH model.
Thus, flavour deconstruction allows the little hierarchy to be more naturally realised in the minimal CH. 

Going in the other direction, we see that the inclusion of a CH solution to the hierarchy problem makes flavour deconstruction more predictive, requiring
    $2M_{W_R}^2 < M_\rho^2$
to avoid flipping the sign of $m_h^2$. 
Thus $v_{23}$ must be sufficiently light to not spoil electroweak symmetry breaking, but the experimental bounds are at the TeV scale (Table~\ref{tab:exp}), so the parameter space is squeezed. Finally, we remark that while the model requires a vector-like fermion with mass $M_\Psi \sim v_{23} (y_3/y_2) \sim 100$ TeV, thanks to the composite dynamics this no longer gives a radiative contribution to $m_h$ proportional to the high scale $M_\Psi$. We believe this kind of model, while seemingly complicated in its construction, is indicative of the benefits (regarding naturalness and predictivity) that can be brought by solving the hierarchy problem and flavour puzzle both at the TeV.

The phenomenology of this model resembles that of the minimal CH model with $U(2)$ flavour protection -- such as modified $HWW$ and $HZZ$ couplings, top partner resonances, and shifts in electroweak observables -- combined with the signatures of the deconstructed $U(1)_Y$ summarized in \S \ref{sec:pheno}. While a full phenomenological study is wanting, the following benchmark scenario is viable: large $g_{R,3}\approx 1$; a light top partner $M_T \approx 2$ TeV with heavier spin-1 resonance $M_\rho \approx 10$ TeV; a deconstruction scale $v_{23}\approx 3$ TeV; which together realises a 5\% or so tuning in the Higgs mass.

\section{Conclusion}

The Higgs remains a central motivation for BSM physics at colliders. In the context of solving the hierarchy problem, the importance of flavour cannot be overlooked. Moving from MFV towards $U(2)$-like flavour symmetries can realise a lower new physics scale, as we saw both from SMEFT studies and for the composite Higgs. Such $U(2)$-like models have the ingredients to also solve the flavour puzzle at a low-scale, for example via non-universal gauge interactions. We reviewed a class of models based on `flavour deconstruction', which do not introduce new sources of flavour violation and so can viable close to the TeV (where naturalness moreover compels them to be, due to non-decoupling Higgs mass corrections). Flavour models in this class predict a rich phenomenology with excellent prospects at HL-LHC and FCC-ee. Finally, these lessons suggest it is fruitful to explore flavour non-universal UV models that solve the flavour puzzle and the hierarchy problem together near the TeV, such as (but not only~\cite{Fuentes-Martin:2020bnh,Lizana:2024jby}) the deconstructed CH model we described.

\section*{Acknowledgments}

I am grateful to S. Covone, G. Isidori, and M. Pesut for collaborating on the most recent work~\cite{Covone:2024elw} that informed this talk.

\end{document}